\input harvmac.tex

%\draftmode

%%%%%%%%%%%%%%%%%%%%%%%%%%%  REFERENCES  %%%%%%%%%%%%%%%%%%%%%%%%%%%%%

\lref\HH{
  J.~B.~Hartle and S.~W.~Hawking,
  ``Wave Function Of The Universe,''
  Phys.\ Rev.\ D {\bf 28}, 2960 (1983).}

\lref\ovv{H.~Ooguri, C.~Vafa and E.~Verlinde,
``Hartle-Hawking wave-function for flux compactifications,''
hep-th/0502211.}

\lref\osv{H.~Ooguri, A.~Strominger and C.~Vafa,
``Black hole attractors and the topological string,''
Phys.\ Rev.\ D {\bf 70}, 106007 (2004) [arXiv:hep-th/0405146].}

\lref\gv{D.~Ghoshal and C.~Vafa,
``C = 1 string as the topological theory of the conifold,''
Nucl.\ Phys.\ B {\bf 453}, 121 (1995) [arXiv:hep-th/9506122].}

\lref\jm{
  J.~Maldacena,
  ``Long strings in two dimensional string theory and non-singlets in the
  matrix model,''
  arXiv:hep-th/0503112.}

\lref\baby{
  R.~Dijkgraaf, R.~Gopakumar, H.~Ooguri and C.~Vafa,
  ``Baby universes in string theory,''
  arXiv:hep-th/0504221.}

\lref\KPZ{V.~G.~Knizhnik, A.~M.~Polyakov and A.~B.~Zamolodchikov,
``Fractal Structure Of 2d-Quantum Gravity,''
Mod.\ Phys.\ Lett.\ A {\bf 3}, 819 (1988).}

\lref\GK{D.~J.~Gross and I.~R.~Klebanov,
``S = 1 for c = 1,'' Nucl.\ Phys.\ B {\bf 359}, 3 (1991).}

\lref\Klebanov{I.~R.~Klebanov,
``String theory in two-dimensions,'' hep-th/9108019.}

\lref\DMP{R.~Dijkgraaf, G.~W.~Moore and R.~Plesser,
``The Partition function of 2-D string theory,''
Nucl.\ Phys.\ B {\bf 394}, 356 (1993)
[arXiv:hep-th/9208031].}

\lref\MP{G.~W.~Moore and R.~Plesser,
``Classical scattering in (1+1)-dimensional string theory,''
Phys.\ Rev.\ D {\bf 46}, 1730 (1992) [arXiv:hep-th/9203060].}

\lref\MPR{G.~W.~Moore, M.~R.~Plesser and S.~Ramgoolam,
``Exact S matrix for 2-D string theory,''
Nucl.\ Phys.\ B {\bf 377}, 143 (1992) [arXiv:hep-th/9111035].}

\lref\MS{G.~W.~Moore and N.~Seiberg,
``From loops to fields in 2-D quantum gravity,''
Int.\ J.\ Mod.\ Phys.\ A {\bf 7}, 2601 (1992).}

\lref\DistlerV{J.~Distler and C.~Vafa,
``A Critical Matrix Model At C = 1,''
Mod.\ Phys.\ Lett.\ A {\bf 6}, 259 (1991).}

\lref\Wittenring{E.~Witten,
``Ground ring of two-dimensional string theory,''
Nucl.\ Phys.\ B {\bf 373}, 187 (1992) [arXiv:hep-th/9108004].}

\lref\Verlinde{E.~Verlinde,
``Attractors and the Holomorphic Anomaly,'' hep-th/0412139.}

\lref\KK{S.~Y.~Alexandrov, V.~A.~Kazakov and I.~K.~Kostov,
``Time-dependent backgrounds of 2D string theory,''
Nucl.\ Phys.\ B {\bf 640} (2002) 119, hep-th/0205079.}

\lref\Malda{J.~Maldacena,
``Non-Gaussian features of primordial fluctuations in single field
inflationary models,'' JHEP {\bf 0305}, 013 (2003) astro-ph/0210603.}

\lref\SYZ{A.~Strominger, S.~T.~Yau, and E.~Zaslow, ``Mirror
Symmetry is T-duality,'' Nucl. Phys. {\bf B479} (1996) 243,
hep-th/9606040.}

\lref\GSVY{B. Greene, A. Shapere, C. Vafa, and S.-T. Yau,
``Stringy Cosmic Strings and Noncompact Calabi-Yau Manifolds,''
Nucl. Phys. {\bf B337} (1990) 1--36.}

\lref\Gross{M.~Gross, ``Topological Mirror Symmetry,''
Invent. Math. {\bf 144} (2001) 75--137;
and ``Special Lagrangian Fibrations I:  Topology,''
in {\it Winter School on Mirror Symmetry, Vector Bundles and
Lagrangian Submanifolds,} C. Vafa and S.-T. Yau, eds.,
AMS/International Press (2001) 65--93.}

\lref\GW{M. Gross and P. M. H. Wilson,
``Large Complex Structure Limits of $K3$ Surfaces,'' math.DG/0008018.}

\lref\SSV{S. Gukov, K. Saraikin, and  C. Vafa, to appear.}

\lref\WZ{E.~Witten and B.~Zwiebach,
``Algebraic structures and differential geometry in $2-D$ string theory,''
Nucl.\ Phys.\ B {\bf 377} (1992) 55, hep-th/9201056.}

\lref\qym{M.~Aganagic, H.~Ooguri, N.~Saulina and C.~Vafa,
``Black holes, q-deformed 2d Yang-Mills, and non-perturbative
topological strings,'' Nucl.\ Phys.\ B {\bf 715} (2005) 304,
hep-th/0411280.}

\lref\anv{M.~Aganagic, A.~Neitzke and C.~Vafa,
``BPS microstates and the open topological string wave function,''
hep-th/0504054.}

\lref\KP{I.~R.~Klebanov and A.~M.~Polyakov,
``Interaction of discrete states in two-dimensional string theory,''
Mod.\ Phys.\ Lett.\ A {\bf 6} (1991) 3273, hep-th/9109032.}

\lref\muva{S.~Mukhi and C.~Vafa,
``Two-dimensional black hole as a topological coset model
of c = 1 string theory,'' Nucl.\ Phys.\ B {\bf 407} (1993) 667,
hep-th/9301083.}

%%%%%%%%%%%%%%%%%%%%%%%%%%%  FIGURES   %%%%%%%%%%%%%%%%%%%%%%%%%%%%%%%

\let\includefigures=\iftrue
\newfam\black
\includefigures
\input epsf
\def\figin{\epsfcheck\figin}\def\figins{\epsfcheck\figins}
\def\epsfcheck{\ifx\epsfbox\UnDeFiNeD
\message{(NO epsf.tex, FIGURES WILL BE IGNORED)}
\gdef\figin##1{\vskip2in}\gdef\figins##1{\hskip.5in}% blank space instead
\else\message{(FIGURES WILL BE INCLUDED)}%
\gdef\figin##1{##1}\gdef\figins##1{##1}\fi}
\def\DefWarn#1{}
\def\figinsert{\goodbreak\midinsert}
\def\ifig#1#2#3{\DefWarn#1\xdef#1{fig.~\the\figno}
\writedef{#1\leftbracket fig.\noexpand~\the\figno}%
\figinsert\figin{\centerline{#3}}\medskip\centerline{\vbox{\baselineskip12pt
\advance\hsize by -1truein\noindent\footnotefont{\bf Fig.~\the\figno:} #2}}
\bigskip\endinsert\global\advance\figno by1}
%%%
\else
\def\ifig#1#2#3{\xdef#1{fig.~\the\figno}
\writedef{#1\leftbracket fig.\noexpand~\the\figno}%
%\figinsert\figin{\centerline{#3}}\medskip\centerline{\vbox{\baselineskip12pt
%\advance\hsize by -1truein\noindent\footnotefont{\bf Fig.~\the\figno:} #2}}
%\bigskip\endinsert
\global\advance\figno by1}
\fi

%%%%%%%%%%%%%%%%%%%%%  Math-style letters   %%%%%%%%%%%%%%%%%%%%%%%%
\font\cmss=cmss10 \font\cmsss=cmss10 at 7pt

\def\IB{\relax\hbox{$\inbar\kern-.3em{\rm B}$}}
\def\IC{\relax\hbox{$\inbar\kern-.3em{\rm C}$}}
\def\IQ{\relax\hbox{$\inbar\kern-.3em{\rm Q}$}}
\def\ID{\relax\hbox{$\inbar\kern-.3em{\rm D}$}}
\def\IE{\relax\hbox{$\inbar\kern-.3em{\rm E}$}}
\def\IF{\relax\hbox{$\inbar\kern-.3em{\rm F}$}}
\def\IG{\relax\hbox{$\inbar\kern-.3em{\rm G}$}}
\def\IGa{\relax\hbox{${\rm I}\kern-.18em\Gamma$}}
\def\IH{\relax{\rm I\kern-.18em H}}
\def\IK{\relax{\rm I\kern-.18em K}}
\def\IL{\relax{\rm I\kern-.18em L}}
\def\IP{\relax{\rm I\kern-.18em P}}
\def\IR{\relax{\rm I\kern-.18em R}}
\def\Z{\relax\ifmmode\mathchoice
{\hbox{\cmss Z\kern-.4em Z}}{\hbox{\cmss Z\kern-.4em Z}}
{\lower.9pt\hbox{\cmsss Z\kern-.4em Z}}
{\lower1.2pt\hbox{\cmsss Z\kern-.4em Z}}\else{\cmss Z\kern-.4em
Z}\fi}

\def\II{\relax{\rm I\kern-.18em I}}

\def\S{{\bf S}}

%%%%%%%%%%%%%%%%%%%%% Calligraphic letters  %%%%%%%%%%%%%%%%%%%%%

\def\CD {{\cal D}}

\def\CF {{\cal F}}

\def\CO {{\cal O}}

\def\CT {{\cal T}}

\def\CW {{\cal W}}

%%%%%%%%%%%%%%%%%%%%%%%%%% Derivatives  %%%%%%%%%%%%%%%%%%%%%%%%

\def\p{\partial}

%%%%%%%%%%%%%%%%%%%% letters with bar %%%%%%%%%%%%%%%%%%%%%%%%%%

\def\hat{\widehat}
\def\bar{\overline}

%%%%%%%%%%%%%%%%%%%%%%%%%%% Math symbols %%%%%%%%%%%%%%%%%%%%%%%

\def\p{\partial}

\def\inbar{\,\vrule height1.5ex width.4pt depth0pt}
\def\r{{\rm Re}}

%%%%%%%%%%%%%%%%%%%   Greek letters %%%%%%%%%%%%%%%%%%%

\def\d{\delta}
\def\e{\epsilon}

\def\m{\mu}

\def\th{\theta}

\def\bar{\overline}

\def\IH{{\bf H}}

%%%%%%%%%%%%%%%%%%%%%%%%%%%%%%%%%%%%%%%%%%%%%%%%%%%%%%%%%%%%%%%%%%%%%%%
%%%%%%%%%%%%%%%%%%% TITLE PAGE  %%%%%%%%%%%%%%%%%%%%%%%%%%%%%%%%

\Title{\vbox{\baselineskip11pt\hbox{hep-th/0505204}
\hbox{HUTP-05/A024}
\hbox{ITEP-TH-37/05}
}}
{\vbox{
\centerline{A Stringy Wave Function for an $S^3$ Cosmology}
}}
\centerline{
Sergei Gukov\footnote{$^{\dagger}$}{On leave from: ITEP, Moscow, 117259, Russia and
L.D.Landau ITP, Moscow, 119334, Russia},
Kirill Saraikin$^{\dagger}$
and  Cumrun Vafa}
\medskip
\medskip
\medskip
%\medskip
\vskip 8pt

\centerline{ \it
Jefferson Physical Laboratory, Harvard University, Cambridge, MA
02138, USA}
\medskip
\medskip
%\medskip
\medskip
%\bigskip
\noindent

Using the recent observations of the relation between
Hartle-Hawking wave function and topological string partition function,
we propose a wave function for scalar metric
 fluctuations on ${\bf S}^3$ embedded in a Calabi-Yau.  This problem
maps to a study of non-critical bosonic string propagating on
a circle at the self-dual radius.   This can be viewed
as a stringy toy model for a quantum cosmology.
\medskip
\Date{May 2005}

%%%%%%%%%%%%%%%%%%%%%%%%%%%%%%%%%%%%%%%%%%%%%%%%%%%%%%%%%%%%%%%%%%%%%%%

\newsec{Introduction}

The notion of the wave function of the universe,
in the mini-superspace description a la Hartle-Hawking \HH,
has recently been made precise in the context of a certain
class of string compactifications \ovv.
In particular, this work provided an explanation
for the appearance of a topological string wave function
in the conjecture of \osv\ relating
the entropy of certain extremal 4d black holes
with the topological string wave function.
It is natural to ask if we can extend this picture to obtain a more realistic
quantum cosmology within string theory.  The aim of the present
paper is to take a modest step in this direction.

The basic setup in \ovv\ was flux compactifications of type II
string theory on a Calabi-Yau three-fold times $\S^2\times \S^1$, and
providing a wave function on moduli space of Calabi-Yau and the overall
size of $\S^2$.  In particular for a given choice of flux in the
Calabi-Yau, labeled by magnetic and electric fluxes $(P^I,Q_I)$, we have
a wave function, $\psi_{P,Q}(\Phi^I)$, depending on (real) moduli of Calabi-Yau.
This wave function is peaked at the attractor values of the moduli
of the Calabi-Yau. Also, in general,
this wave function depends only on the BPS subspace of the field
configurations which thus yields a rather limited information about the
full Calabi-Yau wave function.  One would like to have a wave function
which depends on more local data of the Calabi-Yau geometry, rather than just global moduli.
This may seem to be in contradiction with the requirement that
the data depends only on BPS quantities.
However this need not be the case, as we will now explain.

For concreteness, let us take type IIB superstring compactified
on a Calabi-Yau three-fold and consider the shape of a particular
special Lagrangian 3-cycle $L$ inside the Calabi-Yau.
For instance, this may be a natural setup for a toy model
of our universe obtained by wrapping some D-branes on $L$.
In this setup, the question about the wave function
as a function of the shape of $L$ translates
into the wave function for our universe.
In general, varying the moduli of Calabi-Yau will induce changes
in the shape of $L$. So, at least we have a wave function on
a subset of the moduli of $L$.
More precisely, since Calabi-Yau space has a 3-form which
coincides with the volume form on special Lagrangian
submanifolds, we are effectively asking about a wave function
for some subset of local volume fluctuations on $L$.
On the other hand, since the issue is local, we can consider
a local model of Calabi-Yau near $L$, which is given by $T^*L$.
In this context, global aspects of Calabi-Yau will not provide
any obstruction in arbitrary local deformations of the shape of $L$.
We could thus write a wave function which is a function
of {\it arbitrary local volume fluctuations of} $L$.
In particular, if we know how to compute topological string wave
function on $T^*L$ we will be able to write the full wave function
for arbitrary local volume fluctuations (scalar metric perturbations) of $L$.

A particularly interesting choice of $L$ is $L=\S^3$.
Not only is this the most natural choice in the context
of quantum cosmology, but luckily it also turns out to be
the case already well studied in topological string theory:
As is well known, the topological $B$ model on the conifold
$T^*\S^3$ gets mapped to non-critical bosonic string
propagating on a circle of self-dual radius \gv.
Hence, we can use the results on the $c=1$ non-critical
bosonic string theory to write a wave function for
scalar metric fluctuations of the $\S^3$.
This is the main goal of the present paper.
We will show how the known results of the non-critical
bosonic strings can be used to yield arbitrary 2-point fluctuations.
In particular we find the following result
\eqn\badnews{
\langle \phi_{\vec k}  \phi_{- \vec k} \rangle
\sim {g_s^2 } |\vec k|
}
where $\phi_{\vec k} $ denotes the Fourier modes
of the conformal rescalings of the metric,
which differs from the scale invariant spectrum
in the standard cosmology:
\eqn\goodnews{
\langle \phi_{\vec k}  \phi_{- \vec k} \rangle
\sim  |\vec k|^{-3}
}
One can also in principle compute arbitrary $n$-point fluctuations.
However, in general, for this one would need to know arbitrary momentum
and winding correlation functions of the non-critical bosonic
string which are not yet available (see, however, the recent work \jm).
Nevertheless, from the known results about the correlation functions
of the momentum modes of $c=1$ string theory we can obtain arbitrary n-point fluctuations
for scalar fluctuations on a large circle $\S^1 \subset \S^3$.

The organization of this paper is as follows:
In section 2, we review the notion of the wave function
for topological strings and its relation to the wave
function for moduli of a Calabi-Yau in flux compactifications \ovv.
In section 3, we review non-critical bosonic string theory on a circle
of self-dual radius and its relation to the topological B model on $T^*\S^3$.
In section 4, we use these relations to compute the wave function
for local volume fluctuations on $\S^3$ and compute some
$n$-point correlation functions.
We also discuss some possible toy model cosmologies based on $\S^3$.
Finally, in section 5, we end with the discussion of some open questions and
directions for future research.

%%%%%%%%%%%%%%%%%%%%%%%%%%%%%%%%%%%%%%%%%%%%%%%%%%%%%%%%%%%%%%%%%%%%%%%

\newsec{Stringy Hartle-Hawking Wave Function}

In this section we briefly review the work of \ovv.
Consider a flux compactification of type IIB string
on a Calabi-Yau space $M$ times $\S^2 \times \S^1$,
with a 5-form field strength flux threading through
$\S^2$ and a 3-cycle of $M$.
We choose a canonical symplectic basis for the three cycles on $M$,
denoted by $A_I, B^J$. In this basis, the magnetic/electric flux
can be denoted by $(P^I,Q_J)$.
The wave function of the ``universe'' in the mini-superspace
will be a function of the moduli of $M$ and the sizes of $\S^2$ and $\S^1$.
It turns out that it does not depend on the size\foot{If we change
the boundary conditions on the fermions to be anti-periodic,
then the wave function does depend on the radius of $\S^1$ 
and its norm increases as the value of the supersymmetry breaking
parameter increases \ref\dv{R. Dijkgraaf and C. Vafa, to appear.}.} of the $\S^1$
and its dependence on the size of $\S^2$ can be recast by writing
the wave function in terms of the projective coordinate
on the moduli space of $M$.

The moduli space of a Calabi-Yau is naturally parameterized by
the periods of the holomorphic 3-form $\Omega$ on the 3-cycles.
In particular, if we denote the periods by
\eqn\aperiods{\int_{A_I} \Omega =X^I}
\eqn\bperiods{\int_{B^J} \Omega =F_J}
we can use the $X^I$ as projective coordinates on the moduli space of the Calabi-Yau
(in particular special geometry implies that $F_J$ is determined in terms of $X^I$
as gradients of the prepotential $\CF_0$, i.e. $F_J=\partial_J \CF_0(X^I)$).
However, as observed in \ovv, $X^I$  and ${\overline X}^I$ do not commute
in the BPS mini-superspace.
Therefore, to write the wave function we have to choose a commuting subspace.
A natural such choice is to parameterize this subspace by either real
or imaginary part of $X^I$. Let us call these variables $\Phi^I$.
Then, the wave function is given by
\eqn\wavefunction{\psi_{P^I,Q_J}(\Phi^I)
=\psi_{top}(P^I+{i\Phi^I \over \pi})\ {\rm exp}(Q_J\Phi^J/2)}
where
\eqn\osvfla{\psi_{top}(X^I)={\rm exp}(\CF_{top}(X^I))}
is the B-model topological string partition function.
For compact case there are only finite number of moduli,
but for the non-compact case, which we are interested in here,
there are infinitely many moduli and $I$ runs over an infinite set.
It should be understood that this expression for the wave function
is only an asymptotic expansion (see \baby\ for a discussion
of non-perturbative corrections to this).  The overall rescaling of the charges
is identified with the inverse of topological string coupling constant
and we assume it to be large, so that the string expansion is valid.
The wave function is peaked at the attractor value where
\eqn\attrxf{\eqalign{
& \r X^I=P^I, \cr
& \r F_J=Q_J.
}}

We will be interested in a non-compact Calabi-Yau space,
where the same formalism continues to hold
(one can view it, at least formally, as a limit of a compact Calabi-Yau).
In this case, we will have an infinite set of moduli.
This is similar to \refs{\qym,\anv},
where the incorporation of the infinitely many moduli in the non-compact
case was shown to be crucial for reproducing the conjecture of \osv .
Specifically, in this paper we will be considering the conifold, $T^* \S^3$.
In this case, we can turn on a set of fluxes which result in
a round $\S^3$ at the attractor point, and then consider the fluctuations
of the metric captured by the topological string wave function.
Before we proceed to this analysis, let us review some aspects
of the topological strings on the conifold and its relation
to non-critical bosonic strings on a circle of self-dual radius.

%%%%%%%%%%%%%%%%%%%%%%%%%%%%%%%%%%%%%%%%%%%%%%%%%%%%%%%%%%%%%%%%%%%%%%%

\newsec{Topological Strings on the Conifold and Non-critical c=1 String}

In this section we review the B-model topological string on
the conifold
\eqn\conifold{ xy - zw = \mu}
deformed by the terms of the form $\epsilon (x,y,z,w)$.
The canonical compact 3-cycle of the conifold is $\S^3$.
If we  rewrite \conifold\ as
\eqn\conifoldx{ x_1^2 + x_2^2 + x_3^2 + x_4^2 = \mu }
using appropriate change of the variables the real slice is
exactly this $\S^3$ with  the radius equal to  $\sqrt{ {\rm Re} \mu}$.

Let us recall that these deformations are in 1-to-1 correspondence
with spin $(j,j)$ representations of the
$SO(4) \cong SU(2) \times SU(2)$ symmetry group
\refs{\Wittenring,\WZ,\KP}.
Here, the variables $x_i$ transform in the $(\half, \half)$
representation of the $SU(2) \times SU(2)$.
Thus, we can write $x_i$ as $x^{AA'}$
where $A,A'=1,2$ are the spinor indices.
In these notations, infinitesimal deformations
of the hypersurface \conifoldx\ can be represented
by monomials of the form
\eqn\tmonomial{
\e(x) =  t_{A_1 A_2 \ldots A_n ; A_1' A_2' \ldots A_n'}
x^{A_1 A_1'} x^{A_2 A_2'} \ldots x^{A_n A_n'} }
where the deformation parameters
$t_{A_1 A_2 \ldots A_n ; A_1' A_2' \ldots A_n'}$
are completely symmetric in all $A_i$ and all $A_i'$:
\eqn\fedconifoldx{ x_1^2 + x_2^2 + x_3^2 + x_4^2 = \mu + \e(x)
}
We shall label a generic deformation of the form \tmonomial\
by its quantum numbers:
\eqn\tgroup{
 t_{A_1 A_2 \ldots A_n ; A_1' A_2' \ldots A_n'} \to \ t_{\vert j,j; m,m' \rangle}.
}
where $j=n/2$ and
\eqn\mmprime{\eqalign{
& m = \sum_{i=1}^n (A_i - 3/2) \cr
& m' = \sum_{i=1}^n (A_i' - 3/2)
}}

As is well known \refs{\gv,\DistlerV,\muva}, topological B-model partition function
of the conifold $Z_{{\rm top}}$ considered as a function of the
deformation parameters \tgroup\ can be
identified with the partition function of the  $c=1$
non-critical bosonic string
theory at the self-dual radius
\eqn\ztopzcone{ Z_{{\rm top}} (t) = Z_{c=1} (t) }

The partition function \ztopzcone\ of the $c=1$ theory
is a generating functional for all correlation functions
and has a natural genus expansion in the
string coupling constant $g_s$.
{}From the point of view of the conformal algebra
the equation \conifold\ describes
a relation among four generators of the ground ring \Wittenring,
and ``cosmological constant''  $\mu$ is interpreted as the conifold deformation
parameter.    Moreover the $SU(2)\times SU(2)$ used in classifying
deformation parameters of the conifold get identified with the $SU(2)\times
SU(2)$ symmetry of the conformal theory of $c=1$ at the self-dual radius.
Turning on only the momentum modes leads to deformations
which depend on two of the parameters $\epsilon(x,y)$, whereas turning
on the winding modes corresponds to deformations of the other two variables
$\epsilon(z,w)$.  Turning on all modes corresponds to an arbitrary
deformation of the conifold $\epsilon(x,y,z,w)$, captured by \tmonomial .

The most well studied part of the amplitudes of $c=1$ involves
turning on momentum modes only.  This corresponds to deformation
\eqn\tttconifold{
xy - zw   = \mu + \sum_{n>0}\left( t_n x^n + t_{-n} y^n\right) + \ldots }
Here the dots stand for the terms of higher order in $t_n$ which are
only a function of $x$ and $y$.
The deformations $t_n$, associated with momentum $n$ states,
have the $SU(2)\times SU(2)$ quantum numbers
\eqn\tnstates{
t_n \quad\quad \longleftrightarrow \quad\quad
\vert {|n| \over 2} , {|n| \over 2} ;
{n \over 2} , {n \over 2} \rangle
}
where $n$ runs over all integers.

The partition function \ztopzcone\ for this subset of deformations
is equal to the $\tau$-function of the Toda hierarchy.
In particular, it depends on infinite set of couplings
which are sources for the {\it amputated} tachyon
modes\foot{
Here, $\CT_n = {\Gamma (|n|) \over \Gamma (-|n|)} T_n$ where
$T_n = \int d^2 \sigma e^{(2-|n|)\phi/\sqrt{2}} e^{inX/\sqrt{2}}$
is the standard tachyon vertex. (We follow the conventions
in \refs{\MPR,\DMP}, which are slightly different from
the conventions used in \refs{\GK,\Klebanov}.)
For $n \in \Z$, the vertex operator $\CT_n$ is
a linear combination of a ``special state'' and
the tachyon vertex \refs{\MS,\MPR}.}
\eqn\genfncnf{
\langle
\CT_{n_1} \ldots \CT_{n_k} \rangle
= {\p \over \p t_{n_1}} \ldots {\p \over \p t_{n_k}}
\CF_{c=1} (t) \vert_{t=0}
}
where on the left hand side we have connected amplitudes.
The conservation of momentum implies that the sum
of $n$'s in each non-vanishing amplitude should be equal to zero,
\eqn\momvanishing{
\langle \CT_{n_1} \CT_{n_2} \ldots \CT_{n_k} \rangle
= 0 \quad {\rm unless} \quad \sum_{i=1}^{k} n_i =0
}

The tachyon correlators \genfncnf\ can be computed using
the $\CW_{1 + \infty}$ recursion relations of 2D string theory \DMP.
For example, for genus 0 amplitudes we have
\eqn\ttexamples{\eqalign{
& \langle \CT_{n} \CT_{-n} \rangle = - {\mu^{|n|}\over g_s^2}{1 \over |n|} \cr
& \langle \CT_{n} \CT_{n_1} \CT_{n_2} \rangle = {1 \over g_s^2} \mu^{\half (|n|+|n_2|+|n_3|)-1} \cr
& \langle \CT_{n_1} \CT_{n_2} \CT_{n_3} \CT_{n_4} \rangle =
{1 \over g_s^2} \mu^{\half (|n_1|+|n_2|+|n_3|+|n_4|)-2} (1-{\rm max} \{ |n_i|\}) \cr
& \ldots }}
The genus expansion of the free energy has the form
\eqn\fgenus{ \CF_{c=1} = \sum_{g=0}^{\infty}
\Big( {\mu \over g_s} \Big)^{2-2g} \CF_g (t) }
This is a good expansion in the regime $\mu >> g_s$.
Usually, it is convenient to absorb the string coupling constant
in the definition of $\mu$, and make a suitable redefinition of
the $t_n$'s. However, in our case it is convenient {\it not} to do
this; the advantage is that $t_n$'s appear in the deformed
conifold equation \tttconifold\ without any extra factors.

It is instructive to note that, in this set of conventions,
all the parameters $\mu$, $g_s$, and $t_n$ are dimensionful:
\eqn\dimensions{\eqalign{
& \mu \sim [{\rm length}]^2 \cr
& g_s \sim [{\rm length}]^2 \cr
& t_n \sim [{\rm length}]^{2-|n|}
}}
In particular, the ratio $\big( {\mu \over g_s} \big)$ is
dimensionless, and $t_n$ has the same dimension as $\mu^{1-{|n|
\over 2}}$. Since the genus-$g$ term in the free energy \fgenus\
should be independent of $g_s$, it follows that $\CF_g (t)$
depends on $t_n$ only via the combination $t_n \mu^{{|n| \over
2}-1}$. This is consistent with the fact that when all the $t_n$'s
are zero $\CF_g$ is just a number
\eqn\fzerot{
\CF_g (t_n=0) = {(-1)^{g+1} B_{2g} \over 2g (2g-2)} \quad,\quad g>1
}
In general, $\CF_g (t)$ has the following structure \MPR, \Klebanov
\eqn\fgstructure{
\CF_g (t) = \sum_m
%\Big( t_{n_i} \mu^{{|n_i| \over 2}-1} \Big)^m  P_{m} (n_i)
P^{m}_g (n_i) \prod_{i=1}^m t_{n_i} \mu^{{|n_i| \over 2}-1}
}
where $P^{m}_g (n_i)$ is a polynomial in the momenta $n_i$
of fixed degree depending on $m$ and $g$.
For example, ${\rm deg} P^2_g (n_i) = 4g-1$ and
\eqn\prpone{P^2_1(n) = {1\over 24} (|n|-1)(n^2-|n|-1)
}
Also, $P^m_0 (n_i)$ is a linear polynomial and for $m>2$ is given by
\eqn\kvertex{\eqalign{
P^m_0 (n_i) = (-1)^{m-1} {\mu^{m-2} \over m!}
\Big( \psi_{m-2} + {{\rm max}
\{|n_i|\}\over 2} \sum_{r=1}^{m-3} {(m-2)! \over r! (m-2-r)!}
\psi_{m-2-r} \psi_r\Big),
}}
where $\psi_r := \big({ d \over d\mu} \Big)^r \log \mu$.
Notice that these expressions for $P^3_0(n_i)$
and $P^4_0(n_i)$ agree with \ttexamples.
Thus, the leading genus zero terms have the following form
\eqn\fleading{
\CF_{c=1} = -{1 \over g_s^2} \sum_{n>0} {1 \over n} \mu^n t_n t_{-n}
+ {1\over 3! g_s^2} \sum_{n_1 + n_2 + n_3 = 0}
\mu^{\half (|n_1|+|n_2|+|n_3|)-1} t_{n_1} t_{n_2} t_{n_3} + \ldots
}
It is easy to check that all the terms in this formula
scale as $\mu^2$. It also leads to the tachyon correlation
functions \genfncnf\ consistent with the KPZ scaling \KPZ\
(see also \refs{\GK,\Klebanov }):
\eqn\kpzcorr{
\langle \CT_{n_1} \CT_{n_2} \ldots \CT_{n_k} \rangle_g
\sim \mu^{2(1-g) - k + \half \sum_{i=1}^k |n_i| }
}
%

%%%%%%%%%%%%%%%%%%%%%%%%%%%%%%%%%%%%%%%%%%%%%%%%%%%%%%%%%%

\newsec{Local Volume Form Fluctuations on $\S^3$}

We are interested in making a toy model of quantum cosmology.
In this regard we are interested in the quantum metric fluctuations
of an $\S^3$ inside the Calabi-Yau.  More precisely, we take
the nine-dimensional spatial geometry as
$$M^9= \S^1\times \S^2\times T^* \S^3$$
and study the fluctuations of the metric
on $\S^3\subset T^*\S^3$.
Usually we view the Calabi-Yau scales as much smaller than
the macroscopic scales $\S^1$ and $\S^2$, but nothing in
the formalism of Hartle-Hawking wave function prevents us from
considering larger Calabi-Yau.  In particular we will be assuming
that $\S^3$ has a very large macroscopic size,
which we wish to identify with our observed universe.
One may view our world, in this toy model,
as for example coming from branes wrapped over this $\S^3$,
as we will discuss later in this section.
For the purposes of this section we assume we have certain
fluxes turned on, such that the classically preferred geometry
for $\S^3$ is a large, round metric, and we study what kind
of fluctuations are implied away from this round metric,
in the context of Hartle-Hawking wave function in string theory.
We ask, for example, if the metric fluctuation spectrum implied
by this wave function is scale invariant?

{}From the point of view of the flux compactification
considered in section 2 we
set all electric fluxes to zero
and turn on only one magnetic flux:
\eqn\conifoldflux{\eqalign{
Q_I=0, \cr
P^{I\not =0}=0, \cr
P^0 =N
}}
Then the value of $\mu$ is fixed by the attractor mechanism:
\eqn\muvalue{{\rm Re} \mu= {1\over 2} N g_s}
We will fix $N \gg 1$ leading to ${\rm Re}\mu/g_s \gg 1$.
Note that in this limit the topological string partition
function has a well defined perturbative expansion.

\ifig\fluctuatingsphere{Fluctuations of the $\S^3$
inside a Calabi-Yau.}
{\epsfxsize2.2in\epsfbox{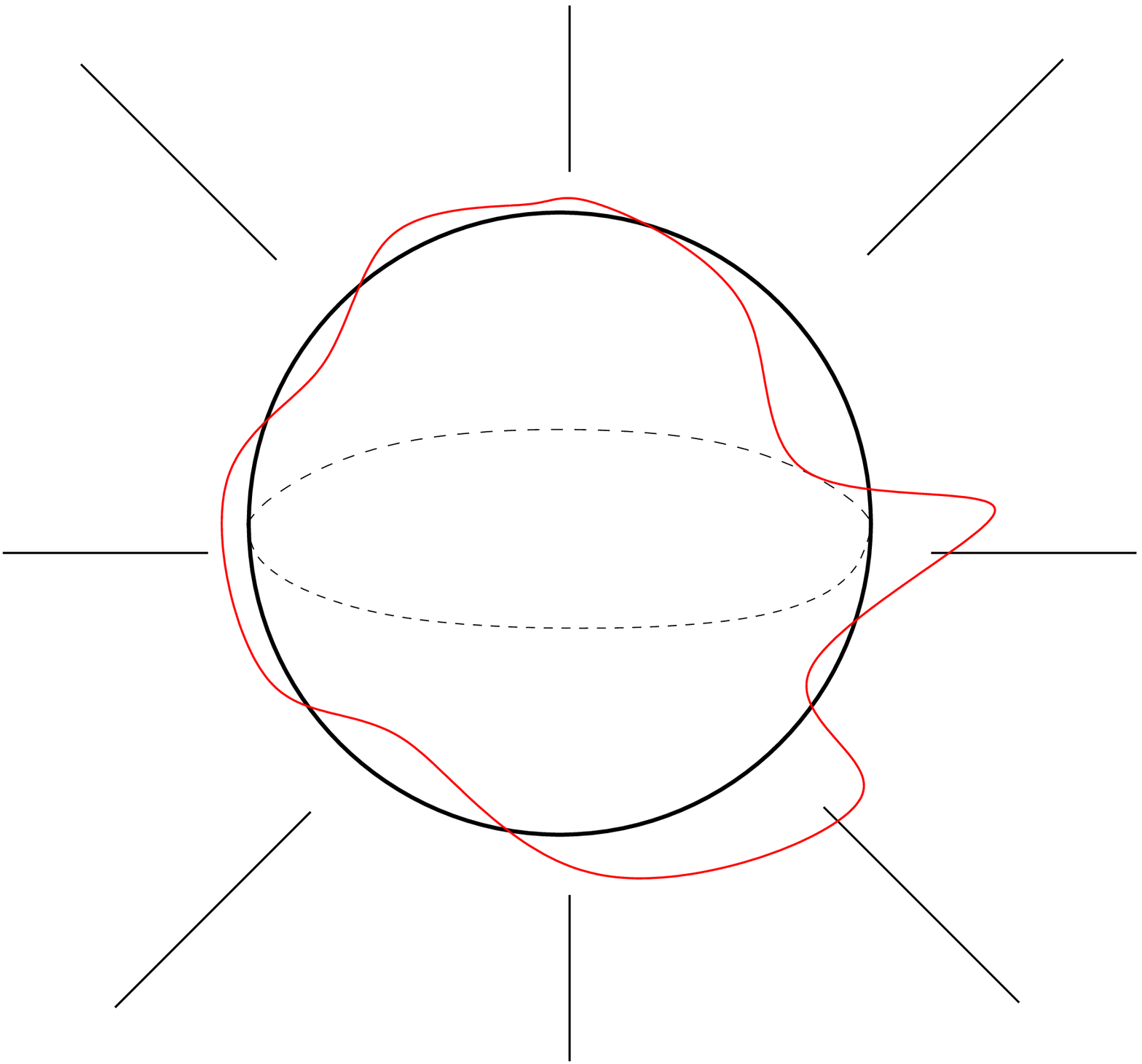}}

Now, let us consider a 3-sphere, $\S^3$, defined by
the real values of the $x_i$, which satisfy eq.\conifoldx.
For non-zero (real) value of the deformation parameter
$\mu$ and zero values of the $t$'s, the induced metric
on the $\S^3$ is the standard round metric, $g_{\mu \nu}^{(0)}$.
The fluctuations of the moduli, $\delta t$, lead to
perturbations of the Calabi-Yau metric on the conifold \conifoldx,
and thus to perturbations of the metric, $g_{\mu \nu}$,
induced on the 3-sphere
\eqn\dtleadsdg{ \delta t \longrightarrow \delta g_{\mu \nu} }
In the topological B-model, the theory depends {\it only} on
the complex structure deformations.  This in particular means
that not all deformations of the metric are observable.  However,
we recall that in the B-model the fundamental field is the holomorphic 3-form $\Omega$
and its variations.  Moreover, on a special Lagrangian submanifold the volume form
coincides with the restriction of a real form of $\Omega $.
In particular, an analog of the scalar fluctuations would be
a fluctuation of the ``conformal factor'' $\phi$, where
\eqn\phidefn{ \Omega = e^{\phi} \Omega_0 }
In particular the field $\phi$ and its fluctuations on a Lagrangian
submanifold would be observable in our wave function induced from
the B-model topological string.  In particular we would be interested
in the fluctuations of the field $\phi$ on the special Lagranigan $\S^3$
inside the conifold.  Before discussing how we do this in more detail,
let us return to what kind of cosmological models would this
question be relevant for.

\subsec{Toy Models of $\S^3$ Cosmology}

So far we have discussed a supersymmetric (morally static) situation,
where we ask the typical local shape of an $\S^3$ inside a Calabi-Yau.
It is natural to ask if we can make a toy cosmology with this data, where
the fluctuations we have studied would be observed as some kind of seed
for inhomogeneity of fluctuations of matter.

In order to do this we need to add a few more ingredients to our story:
First of all, we need to have the observed universe be identified with what
is going on in an $\S^3$.  The most obvious way to accomplish this would be
in the scenario where we identify our world with some number of D3 branes
wrapping $\S^3$.  In this situation the inhomogeneities of the
metric on $\S^3$ will be inherited by the D3 brane observer.
A second ingredient we need to add to our story would be time dependence.
This would necessarily mean going away from the supersymmetric context--an
assumption which has been critical throughout our discussion.
The least intrusive way, would be to have our discussion be applicable
in an adiabatic context where we have a small supersymmetry breaking.
In particular we imagine a situation where time dependence of the fields
which break supersymmetry is sufficiently mild, that we can still trust a mini-superspace
approximation in the supersymmetric sector of the theory.

To be concrete we propose one toy model setup where both of these can in principle
be achieved.  We have started with no D3 branes wrapped around $\S^3$.  In the
context of attractor mechanism this means that
$${\rm Re }\hat \mu =P$$
$${\rm Re}{i\over 2\pi}\hat \mu {\rm log}{\hat \mu\over \Lambda}=Q=0$$
(where $\hat \mu =2\mu/g_s$)
which we realize by taking $\hat \mu$ to be real and equal to $P$ and $\Lambda$ to be real.
The value of $\Lambda$ is set by the data at infinity of the conifold.  Let us write
$$\Lambda =\Lambda_0 {\rm exp}(i\varphi)$$
and imagine making $\Lambda$ time dependent by taking a time dependent $\varphi(t)$.
This can be viewed as a ``time dependent axion field'' induced from
data at infinity.
This leads to creation of flux corresponding to D3 brane
wrapping $\S^3$ as is clear from the attractor mechanism.
Indeed each time $\varphi$ goes through $2\pi$ the number
of D3 branes wrapping $\S^3$ increases by $P$ units.
This in turn can nucleate the corresponding D3 branes.

To bring in dynamics leading to evolution of radius
of $\S^3$ we can imagine the following possibilities:
Make the magnetic charge $P$ time dependent by bringing
in branes from infinity in the same class (or perhaps
by the magnetic brane leaving and annihilating other
magnetic anti-branes, leading to shrinking $\S^3$).
This can in principle be done in an adiabatic way,
thus making our story consistent with a slight time dependent $\mu$.
Another possibility which would be less under control would be
to inject some energy on the D3 branes.
It is likely that this leads to some interesting
evolution for $\S^3$, though this needs to be studied.
In particular, this can be accomplished by making the $\varphi(t)$
undergo partial unwinding motion.
In this way we would create some number of anti-D3 branes
which would annihilate some of the D3 branes.
It is interesting to study what kind of cosmology this would lead to.
Keeping this toy model motivation in mind  we now return to
the study of volume fluctuations in the supersymmetric model.

\subsec{Setup for Computation of Volume Fluctuations}

In principle if we know the full amplitudes of the $c=1$ theory at
the self-dual radius we can compute all correlation functions of $\phi$.
However the full amplitudes for $c=1$ are not currently known
(see \jm\ for recent work in this direction).
We thus focus on the amplitudes which are known,
which include the momentum mode correlations,
as discussed in section 3.  For two point correlation functions,
as we will note below, the general amplitudes can be read off from
this subspace of deformations, due to $SU(2)\times SU(2)$ symmetry of $\S^3$.

The momentum induced deformation of the 3-sphere in the conifold
geometry \tttconifold\ is obtained by specializing to a real
three-dimensional submanifold, described by the equation
\eqn\deformedconex{\eqalign{
p  + x_3^2 + x_4^2 = \mu + \e(p, \theta)}}
where $x_3, x_4$ are real and without loss of generality,
$\mu$ is assumed to be real
and
\eqn\xyviaptheta{\eqalign{
x & = p^{1/2} e^{i \th} \cr
y & = p^{1/2} e^{- i \th}
}}
In these variables, the restriction of the holomorphic 3-form $\Omega$
to the hypersurface \deformedconex , which is the volume form on it,
is given by
\eqn\deformthreeform{\eqalign{
\Omega = {{dx_3 dx_4 d\th} \over {1- \p_p \e(p,\th) }},
}}
In particular, in the linear approximation
\eqn\deformdeterminant{\eqalign{
 \e(p, \th) \approx \r \sum_{n\not=0}p^{|n|/2} e^{i n \th} t_n
 }}
The fluctuations of the ``conformal factor'' are given by
\eqn\volumefluct{\eqalign{
\phi = \log {{ \Omega} \over {\Omega_0}} =
-\log (1- \p_p \e) = \p_p \e + \half (\p_p \e)^2+ \ldots
}}
Now let us look at the absolute value squared of
the Hartle-Hawking wave function \osvfla. Remember
that relation between the $c=1$ theory at
the self-dual radius and the B-model topological string
on the conifold \ztopzcone\ implies
$\CF_{top}(t) =\CF_{c=1}(t)$. Therefore,
\eqn\psisquared{
|\Psi|^2 = \exp \Big(
-{1 \over g_s^2} \sum_{n>0} {2 \over n} \mu^n \r (t_n t_{-n})
+ {1\over 3 g_s^2}\sum_{n_1 + n_2 + n_3 = 0}  \mu^{\half (n_1+n_2+n_3)-1}
\r (t_{n_1} t_{n_2} t_{n_3}) + \ldots
\Big) }
We are going to use this wave function density to evaluate
correlation functions of the form:
\eqn\nptcorr{
\langle \phi_1 \phi_2 \ldots \phi_n  \rangle
= {\int \CD t (\phi_1 \phi_2 \ldots \phi_n) |\Psi (t)|^2
\over \int \CD t |\Psi (t)|^2}
}
where $\phi_k = \phi (p=\mu,\th_k)$ is the conformal
factor at a point on the ``large circle'' of the $\S^3$,
defined by eq. \deformedconex\ with $x_3 = x_4 =0$.
As we already noted, a computation
of more general correlations functions (where $\phi_k$
are in general position on the $\S^3$) would require
the information about the correlation functions of both
momentum and winding modes of the $c=1$ model.
The reason for this is that $p \ne \mu$ implies
$x_3^2 + x_4^2 \ne 0$ and, therefore, leads to
generic deformations $\epsilon (p,\th,x_3,x_4)$
in \deformedconex.
Since from now on we will always consider only
the correlation functions of the conformal factor
on the large circle, $p=\mu$, we shall often write
$\phi(\mu,\th) = \phi(\th)$.  We will now turn to the two
point function for which the momentum correlation functions
are sufficient to yield the general correlation function
due to $SU(2)\times SU(2)$ symmetry.

\ifig\nptsonthecircle{$n$ points on the large circle of the $\S^3$.}
{\epsfxsize2.2in\epsfbox{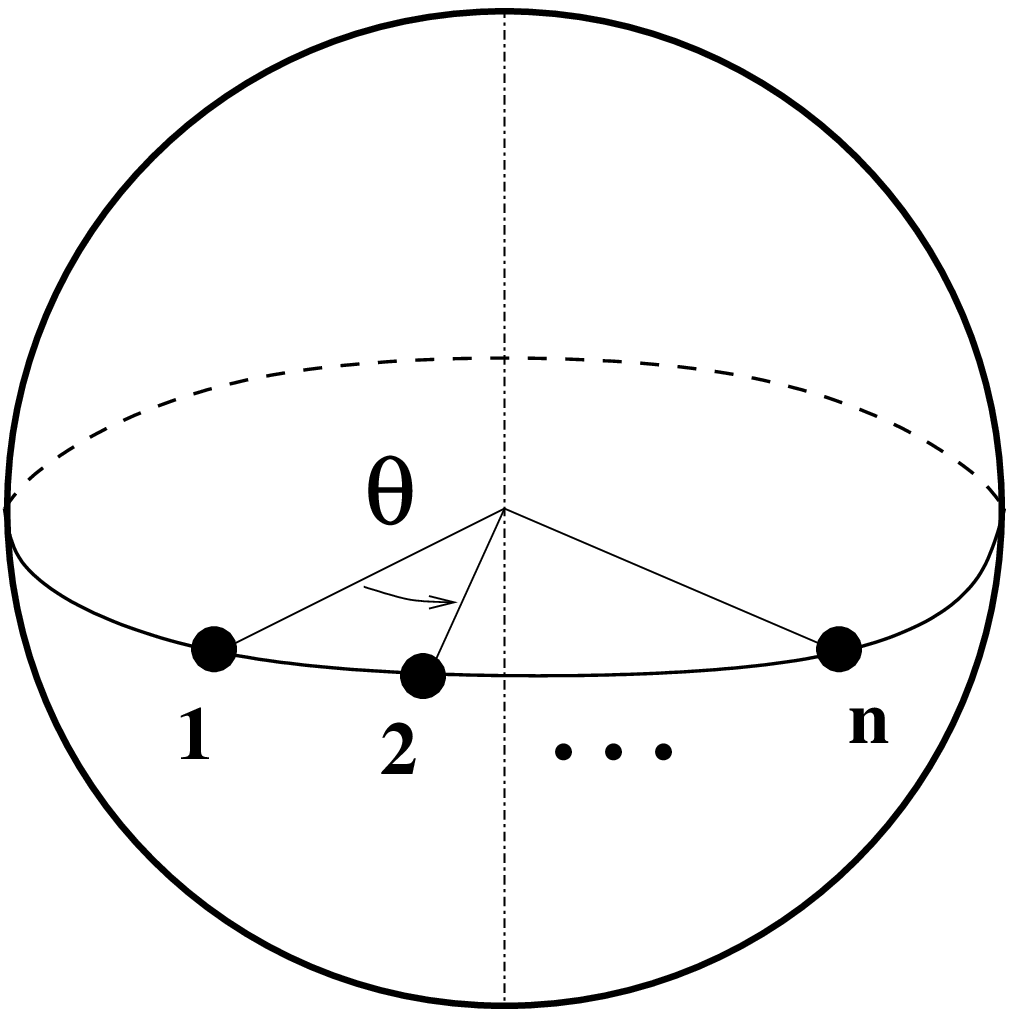}}

%%%%%%%%%%%%%%%%%%%%%%%%%%%%%%%%%%%%%%%%%%%%%%%%%%%%%%%%%

\subsec{2-point Function at Tree Level}

We start with evaluating a two point correlation function:
$$
\langle \phi_1 \phi_2 \rangle
$$
Because of the $SO(4)$ symmetry of the $\S^3$, one can
always assume that $\phi_1$ and $\phi_2$ are evaluated
at two points on the large circle $p=\mu$.
To the leading order in $g_s/\mu$, one can keep only
the linear terms in \volumefluct. The contribution of
non-linear terms in \volumefluct\ is suppressed by $g_s$
and will be discussed later. Thus, in the linear approximation to \volumefluct\
and substituting \deformdeterminant ,
\eqn\twopt{\eqalign{
\langle \phi(\th) \phi(0) \rangle = {1\over 4 \m^2}\sum_{n,m} |n m|  \m^{{|n|+|m|\over 2}}
e^{in\th}\langle t_n t_{m}  \rangle
 }}
where we used the fact that $t_n$ and $t_{-n}$
are complex conjugate after reduction to the 3-sphere.
Similarly, restricting \psisquared\ to the 3-sphere, we get
\eqn\psisphere{\eqalign{
|\Psi_{\S^3}(t)|^2=\exp\left(- {2 \over g_s^2}
\sum\limits_{n>0} { \mu^{|n|} \over |n|}  t_n t_{-n}
+ {1\over 3 g_s^2 \m} \sum\limits_{n_1 + n_2 =- n_3 } \mu^{{|n_1|+|n_2|+|n_3|\over 2}}
t_{n_1} t_{n_2} t_{n_3} + \ldots \right)
 }}
In particular, it gives
\eqn\ttdef{\eqalign{
\langle t_n t_{m}  \rangle = {\int \CD t |\Psi_{\S^3}(t)|^2 t_n t_{m} \over
\int \CD t |\Psi_{\S^3}(t)|^2}=
{|n| \over 2} g_s^2 \mu^{-|n|}  \d_{n+m,0} + \CO(({\m / g_s})^{-|n|-2})
 }}
Evaluating the path integral we treat non-quadratic terms in
\psisphere\ as perturbations. In general, this will give a
highly non-trivial theory with all types of interactions.
However, using scaling properties \dimensions,  one can show
that contribution from the $k$-tuple interaction vertex
is proportional to $(g_s/\m)^{k-2}$ (see discussion below).
Therefore, all loop corrections to the leading term are
suppressed in the limit of large $\S^3$ radius and
small string coupling, $g_s$. As a result, using \ttdef\ we find
\eqn\twoptsum{\eqalign{
\langle \phi(\th) \phi(0) \rangle ={g_s^2 \over 8 \mu^2}\sum_{n} |n|^3 e^{in\th} + \dots
 }}
or, equivalently,
\eqn\twoptmomentum{
\langle \phi_n  \phi_{-n} \rangle = {g_s^2 \over 8 \mu^2} |n|^3 }
where
\eqn\phindef{ \phi_n := {1 \over 2\pi} \int_0^{2 \pi} d \th e^{i \th n} \phi(\th)}
This has to be compared with the usual Fourier transform,
given by a 3-dimensional integral
\eqn\kkktwopt{\eqalign{
\langle \phi_{\vec k}  \phi_{- \vec k} \rangle
& \sim \int d^3 x e^{i \vec k \cdot \vec x} \langle \phi(\vec x)
\phi(0) \rangle \cr
& \sim {g_s^2 } |\vec k|
}}
Note that a scale invariant power spectrum would correspond
to $|k|^{-3}$ fluctuation correlation.  Thus the fluctuation
spectrum we have on $\S^3$ is {\it not} scale invariant.

After performing the summation over $n$ in \twoptmomentum,
we get the 2-point function
\eqn\twoptglobal{\eqalign{
\langle \phi(\th) \phi(0) \rangle
= {g_s^2 \over 32 \mu^2} {\cos\th +2 \over \sin^4 \th /2}
 }}
in the coordinate representation. Although this expression
appears to have a singularity at $\th = 0$, as we explain in
the next section, our approximation cannot be trusted at
large momenta or, equivalently, small $\th < \sqrt{g_s/\mu}$.

%%%%%%%%%%%%%%%%%%%%%%%%%%%%%%%%%%%%%%%%%%%%%%%%%%%%%%%%%%

\subsec{General Structure of $g_s$ Corrections}

There are three sources for the $g_s$ corrections to the 2-point function:
$i)$ one due to higher genus terms in the free energy expansion \fgenus,
$ii)$ corrections due to loops made from the $k$-point vertices (with $k>2$)
in the ``effective action'' $\CF (t)$ as well as due to non-linear
terms in the expansion \volumefluct\ of $\phi $ in terms of $\epsilon$, and $iii)$ corrections
due to non-linear relation between $\epsilon$ and the deformation parameters
$t_n$ induced by the deformations of the geometry \refs{\DMP,\KK}:
\eqn\deformedconexsec{\eqalign{
 x y   = \mu -  x_3^2 - x_4^2 + \sum_{n>0}( t_n x^n + t_{-n} y^n) -
 {1\over 2 \m} \sum_{{ m>0 \atop n>0 }}  t_{n} t_{-m} m   x^{n}y^{m} + \ldots
 }}
It is easy to check that all kinds of corrections
are suppressed by powers of $(g_s/\mu)^2$.
In the case $i)$ this is manifest from the form of \fgenus.
In the case $ii), iii)$, this can be seen in the language
of the Feynman diagrams for the fluctuating fields $t_n$,
that follow from the effective action \psisphere:
\eqn\feynmanrules{\eqalign{
& {\rm propagator:}~~~~~~~~~~~~~~~~~ \half g_s^2 |n| \mu^{-|n|} + \ldots \cr
& k \ge 3 {\rm ~vertex:}~~~~~~~~~~~~
{2 \over g_s^2 k!} \mu^{{|n_1| + \ldots + |n_k| \over 2}+2-k} P_0^k (n_i) + \ldots
}}
Here $P_0^k (n_i)$ is a linear polynomial \kvertex\ in momenta $n_i$,
and the dots stand for higher-genus terms. In particular,
a genus-$g$ contribution comes with an extra factor of $(g_s/\mu)^{2g}$.

In general, we find that the genus-$g$ contribution
(contribution from $g$ loops) to the 2-point function
looks like
\eqn\twoptgenusg{
\langle \phi_n  \phi_{-n} \rangle_g
\sim \Big( {g_s \over \mu} \Big)^{2g+2} |n|^{4g+3}
}
where $\phi_n$ is defined in \phindef.
We can read off the higher genus corrections to the propagator
$\langle t_n t_{-n}\rangle$ from
the quadratic terms in \fgstructure:
\eqn\propgen{
\langle t_n t_{-n}\rangle \big|_{tree} =
{ \half g_s^2 |n| \mu^{-|n|}  \over {
1 + \sum\limits_{g \geq 1} ({g_s / \mu})^{2g} |n| P^2_g (n)}}
}
Notice that  ${\rm deg} P^2_g (n) = 4g-1$ and therefore for large momenta we have an
asymptotic expansion of the form:
\eqn\propas{
\langle t_n t_{-n}\rangle \big|_{tree}
= { \half g_s^2 |n| \mu^{-|n|}  \over {
1 + \sum\limits_{g \geq 1} p_g \Big({n^2 g_s \over \mu}\Big)^{2g} + \ldots  }}
}
where constants $p_g$ are determined by the polynomial $P^2_g (n)$
and dots stand for the terms suppressed by powers of $n$.
Now it is clear that
the good expansion parameter is ${n^2 g_s \over \mu}$ rather than
${g_s \over \mu}$
which means that our approximation is valid only for momenta $n$
small compared to $\mu/g_s$. In other words, we should fix some high-energy
cut-off parameter $\Lambda^2 < \mu/g_s$ and consider only
deformations with momentum number $n< \Lambda$.

Now let us incorporate corrections due to loops
in Feynman diagrams generated by \feynmanrules.
For example, if we take into account genus one
corrections and one-loop corrections we get
the following expression for the propagator:
\eqn\propgenustwo{
{\lower3.0pt \hbox{\epsfxsize1.6in\epsfbox{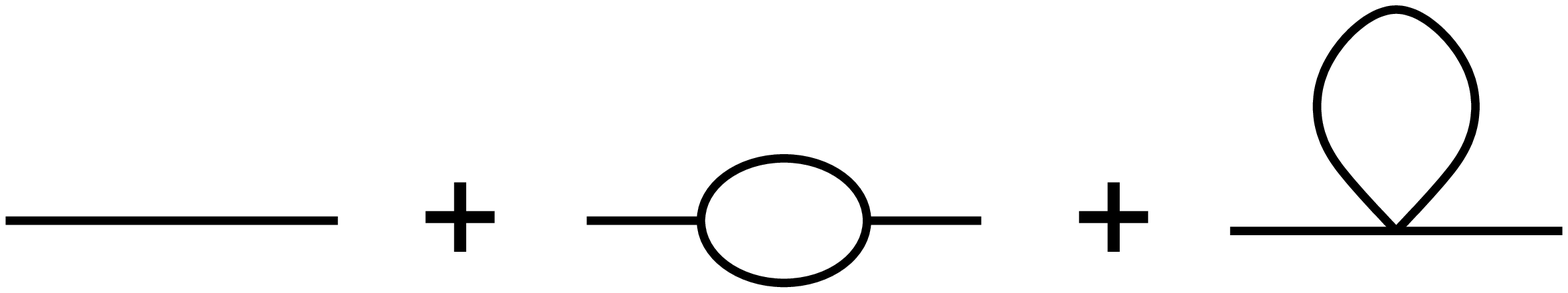}}} + \ldots =
{ \half g_s^2 |n| \mu^{-|n|}  + \ldots  \over {
1 - {|n| \over 24} \Big({g_s \over \mu}\Big)^2 (|n|-1)(n^2 -|n|-1)
+ \ldots }}
}
Notice that due to \twoptgenusg\ $g$-loops corrections dependence on
momenta is similar to the genus $g$ corrections.
Thus, the general structure of the 2-point function is given by:
\eqn\twoptgeneral{
\langle \phi_n  \phi_{-n} \rangle_g
= {g_s^2 \over 8 \mu^2} |n|^3 { 1+ \sum\limits_g b_g
\big( {n \over \Lambda} \big)
\Big( {\Lambda^2  g_s  \over \mu} \Big)^{2g} +\ldots
\over 1+ \sum\limits_g p_g \Big( {n^2 g_s \over \mu} \Big)^{2g} +\ldots }
}
where the polynomials $b_g$ depend on the ratio $n/\Lambda$,
which should be small in order for the perturbation theory to be valid.
%is expected to be valid as long as  $n< \Lambda$.

%%%%%%%%%%%%%%%%%%%%%%%%%%%%%%%%%%%%%%%%%%%%%%%%%%%%%%%%%%

\subsec{$n$-point Function for the Perturbations on the Large Circle of $\S^3$}

Here we briefly discuss the structure of $n$ point function.  Unlike
the 2-point function where we could compute the general
case, for $n$ point functions with the present technology, we can
only compute correlations restricted to taking the fluctuations at points
 on a large circle.
Using the Feynman rules \feynmanrules,
we find that the contribution of a tree Feynman
diagram to a $k$-point function scales as (to avoid cluttering,
we omit polynomials in $n_i$ which do not affect the $g_s$ behavior):
\eqn\tkfncngzero{
\langle t_{n_1} \ldots t_{n_k} \rangle_0 \sim
g_s^{2k-2} \mu^{-{|n_1| + \ldots + |n_k| \over 2}+2-k}
}
Now, let us consider a $g$-loop contribution to
the $k$-point function. As we discussed earlier,
such contributions come from the
vertices with a total of $k+2g$ legs, $k$ of which are connected
by propagators to the external legs of the $k$-point function,
and $2g$ of which are pairwise connected by internal propagators.
Notice that, for the internal momenta $n_j$,
the factors $\mu^{{|n_j| \over 2}}$ cancel out and we get
\eqn\tkfncng{
\langle t_{n_1} \ldots t_{n_k} \rangle_g \sim
g_s^{2k+2g-2} \mu^{-{|n_1| + \ldots + |n_k| \over 2}+2-k-2g}
}
Comparing this expression with \tkfncngzero, we see that
a $g$-loop contribution to the $k$-point correlation
function is suppressed by the same factor $(g_s/\mu)^{2g}$
as the contribution from a genus-$g$ term in the free energy \fgenus.
For the $k$-point function of the fields $\phi_n$ this implies
\eqn\phinkcorr{\eqalign{
\langle \phi_{n_1} \ldots \phi_{n_k} \rangle_g
& \sim \mu^{{|n_1| + \ldots + |n_k| \over 2}-k}
\langle t_{n_1} \ldots t_{n_k} \rangle_g \cr
& \sim \Big( {g_s \over \mu} \Big)^{2k+2g-2}
}}
where in the second line we used \tkfncng.
Notice, this structure is consistent with our
results \twoptmomentum\ for the 2-point function.  For
an example of higher point function we now turn to a discussion
of the leading correction to the 3-point function.

%%%%%%%%%%%%%%%%%%%%%%%%%%%%%%%%%%%%%%%%%%%%%%%%%%%%%%%%%%

\subsec{3-point Function}

Now, let us look more carefully at the structure of the 3-point
function.  Unlike the 2-point function where we studied the general case,
since the topological string amplitudes are not known for arbitrary deformations
of the conifold, we restrict our attention to the ones corresponding
to momentum modes.  This means that we consider 3-point functions where all three points lie on the large circle of the $\S^3$.
To the leading order in $(g_s/\mu)^2$, from \volumefluct\ we find
\eqn\threeptfirst{\eqalign{
\langle \phi (\th_1) \phi (\th_2) \phi (\th_3)\rangle=
{1\over 8 \m^3}\sum_{n,m,l} |n m l|  \m^{{|n|+|m| +|l|\over 2}}
e^{i n\th_1 + i m\th_2 + i l\th_3}\langle t_n t_{m} t_{l} \rangle
 }}
In the momentum representation, this looks like
\eqn\threeptfirstmom{\eqalign{
\langle \phi_n \phi_m \phi_l \rangle=
{1\over 8 \m^3} |n m l| \m^{{|n|+|m| +|l|\over 2}}
\langle t_n t_{m} t_{l} \rangle
}}
According to \psisphere,
\eqn\tthreept{\eqalign{
 \langle t_n t_{m} t_{l} \rangle ={g_s^4 \over 4}  \d_{n+m+l,0} |n m l|
 \m^{-{|n|+|m| +|l|\over 2} - 1}
 }}
which gives
\eqn\threeptcontmomentum{ \langle \phi_n \phi_m \phi_l \rangle
= {1 \over 32} \Big( {g_s \over \mu} \Big)^4 |n m l|^2 \d_{n+m+l,0} }
This is to be compared with the three point function of fluctuations
in the inflationary cosmology \Malda.
%

%%%%%%%%%%%%%%%%%%%%%%%%%%%%%%%%%%%%%%%%%%%%%%%%%%%%%%%%%%

\newsec{Further Directions}

In section 4 we mentioned some potential cosmological models
based on the $\S^3$ in the conifold geometry.
It would be very interesting to develop these ideas further.
It is also natural to ask if in some other non-compact models
we could get a different spectrum of fluctuations.
It is probably true that the local Calabi-Yau fluctuations
that we find are fairly generic, but the global
aspects might be different for different Calabi-Yau.

As for the local fluctuations, it would be interesting to relate
our results to properties of the Kodaira-Spencer theory,
which is the target space theory of topological gravity.
In particular, it should be possible to derive
the 2-point function of the local fluctuations directly from
the relation between the Kodaira-Spencer theory, the B-model,
and its interpretation as the Hartle-Hawking wave function.

Another aspect of the wave function that topological strings lead to
has to do with the fundamental question
of `which' Calabi-Yau space is preferred?
Naively, one would think that it is not possible to answer
this question in our setup, because the probability density
depends on the scale of the holomorphic 3-form,
which in turn corresponds to overall rescaling of the black hole charge.
In particular, the overall rescaling of the black hole charge
does not affect the attractor point on the moduli space of the Calabi-Yau,
but makes the entropy of the black hole arbitrarily large
for {\it any} point on the moduli of Calabi-Yau.
However, it turns out that there is a way to extract information
about which Calabi-Yau is preferred \SSV: We consider a local model
of a Calabi-Yau, very much in line with the example considered in this paper.
In this way we can partially decouple gravity from our wave function
by fixing the period of the holomorphic 3-form on a particular 3-cycle,
$$
\int_{A_0}  \Omega  \sim {1\over g_{s}}
$$
With this constraint, we then look for the local
maxima of the entropy functional,
$$
S= -i {\pi \over 4} \int_{CY} \Omega \wedge \bar  \Omega.
$$
It turns out, that certain singularities of the Calabi-Yau
on the moduli space extremize $S$ subject to the above constraint.
Moreover, in these models, we find a correlation between the sign
of the beta function in the effective four-dimensional field theory
and the second order variation of the functional $S$.
In particular, we find that $S$ is maximized
{\it for asymptotically free theories}!
Further details of these models will appear elsewhere \SSV.

%%%%%%%%%%%%%%%%%%%%%%%%%%%%%%%%%%%%%%%%%%%%%%%%%%%%%%%%%%%%%%%%%%%%%%%

\vskip 30pt

\centerline{\bf Acknowledgments}

We would like to thank N. Arkani-Hamed, R.Dijkgraaf, A. Neitzke,
and H. Ooguri for valuable discussions.
This research was supported in part by NSF grants PHY-0244821
and DMS-0244464. This work was conducted during the period
S.G. served as a Clay Mathematics Institute Long-Term Prize Fellow.
K.S. and S.G. are also supported in part by RFBR grant 04-02-16880.

\listrefs
\end